\documentclass[12pt]{article}
\usepackage{epsfig}
\title{Search for the radiative decay $\eta \to \pi^0 \gamma \gamma$ 
in the SND experiment at VEPP-2M }
\author
{ M.~N.~Achasov, K.~I.~Beloborodov, A.~V.~Berdyugin, A.~V.~Bozhenok,\\
A.~G.~Bogdanchikov, D.~A.~Bukin, S.~V.~Burdin, T.~V.~Dimova, \\ 
A.~A.~Drozdetsky, V.~P.~Druzhinin, D.~I.~Ganyushin, V.~B.~Golubev, \\ 
V.~N.~Ivanchenko, I.~A.~Koop, A.~A.~Korol, S.~V.~Koshuba, \\ 
A.~V.~Otboev, E.~V.~Pakhtusova, A.~A.~Salnikov, V.~V.~Shary, \\
S.~I.~Serednyakov, Yu.~M.~Shatunov, V.~A.~Sidorov, Z.~K.~Silagadze 
\thanks {Corresponding author, e-mail silagadze@inp.nsk.su},\\ 
A.~G.~Skripkin, A.~A.~Valishev, A.~V.~Vasiljev and Yu.~S.~Velikzhanin
\vspace*{3mm} \\
Budker Institute of Nuclear Physics and \\ \vspace*{1mm}
Novosibirsk State University, 630 090, Novosibirsk, Russia }

\date{}

\begin{document}
\large
\maketitle

\begin{abstract}
The $\eta \to \pi^0 \gamma \gamma$ decay was investigated by the SND
detector at VEPP-2M $e^+e^-$ collider in the reaction $e^+e^-\to\phi\to
\eta\gamma$. Here we present the results and some details of this 
study. We report an upper limit (90\% c.l.) 
$Br(\eta \to \pi^0 \gamma \gamma)<8.4\times 10^{-4}$ as our final result.
Our upper limit does not contradict the earlier measurement by GAMS 
spectrometer.
To facilitate future studies a rather detailed review of the problem is also
given. 
\end{abstract}

\newif\iftranslate
\translatefalse

\section{Introduction}
The experimental history of the $\eta \to \pi^0 \gamma \gamma$ decay
begins in 1966 when $1.2~\mathrm{GeV/c}$ momentum $\pi^-$ beam from
CERN proton synchrotron was used to produce $\eta$ mesons on a hydrogen
target via charge exchange reaction
\begin{equation}
\pi^- +p \to \eta +n.
\label{eq:cher} \end{equation}
\noindent $\eta$ meson was tagged by means of neutron time-of-flight 
measurement.
The $\gamma$-ray emission spectrum
from the subsequent $\eta$ decays was measured by a lead 
glass Cherenkov counter. As it was claimed in Ref.\cite{1}, the observed 
$\gamma$-ray spectrum could not be explained by 
$\eta \to 2\gamma$ and $\eta \to 3\pi^0$ decay modes only. An equally 
intensive $\eta \to \pi^0 \gamma \gamma$ decay was required to get
a good approximation of the measured spectrum.

The next experiment carried out in Brookhaven \cite{2} utilized a similar 
technique
except for an order of magnitude higher incident $\pi^-$ beam momentum and
a spark chamber as a $\gamma$-ray detector. No $\eta \to 
\pi^0 \gamma \gamma$ signal was observed in the collected 4$\gamma$ events.

\begin{figure}[htb]
\resizebox{0.95\textwidth}{!}{
  \includegraphics{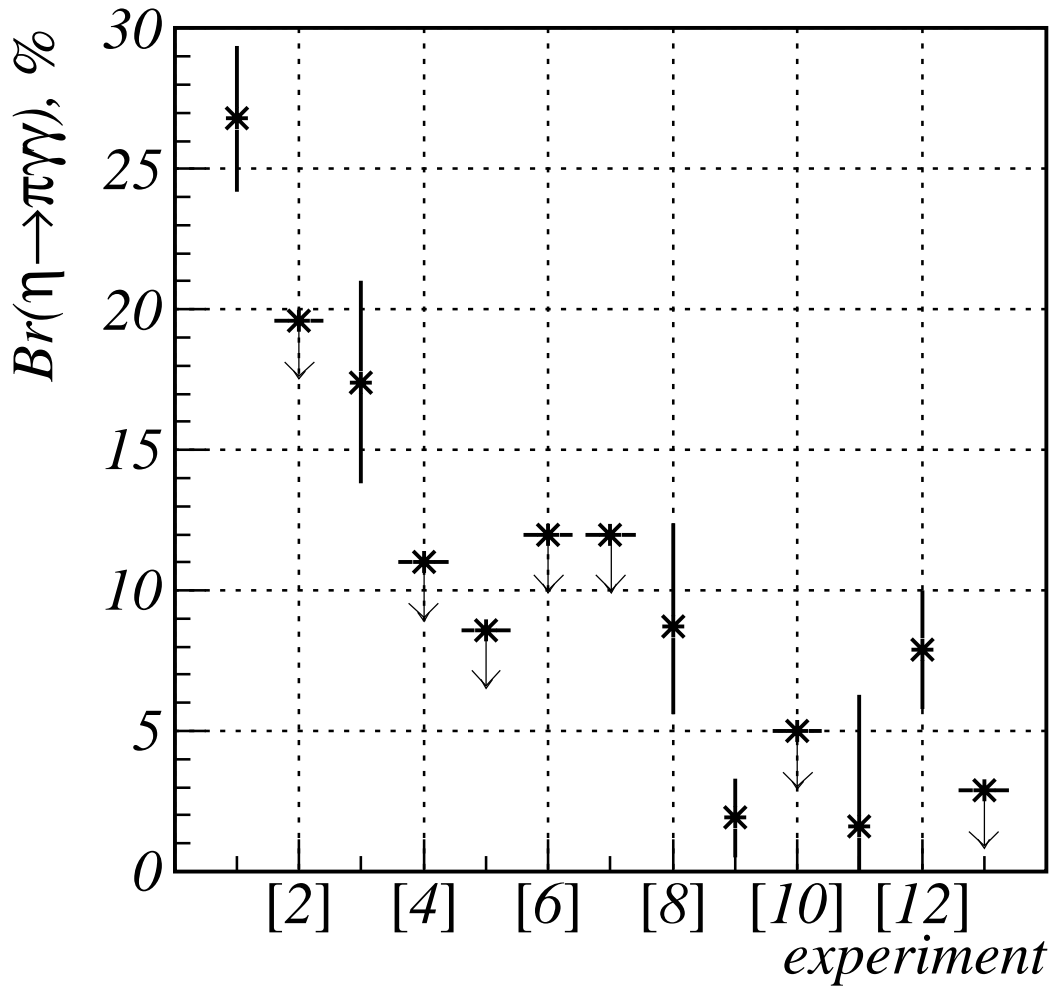}
}
\caption{The history of the $\eta \to \pi^0 \gamma \gamma$ decay 
measurements up to 1980.}
\label{fig:br}       
\end{figure}

That was the beginning of the $\eta \to \pi^0 \gamma \gamma$ decay puzzle
which lasted for the next 15 years. Several other experiments were performed
during this time $[3\div 13]$,
but the situation still remained unclear.
The results of these experiments are shown in Fig.\ref{fig:br}.
For experiments that did not measure this branching ratio directly we
calculated it using their results and current PDG table values for
$\mathrm{Br}(\eta\to 2\gamma)=(39.21\pm 0.34)\%$ and
$\mathrm{Br}(\eta\to\mathrm{neutrals})=(71.5\pm 0.6)\%$.

In many experiments the reaction (\ref{eq:cher}) with rather low
$\pi^-$ momentum $\sim 1~\mathrm{GeV/c}$ was used as a source of
$\eta$ mesons and optical spark chambers as
$\gamma$-ray detectors, as in \cite{2}.
Rather low detection efficiency for photons lead to 
a severe background  from intense $\eta \to 3\pi^0$ decay.
This was clearly demonstrated in
\cite{11}. It is interesting to note, however, that despite the authors'
firm
conclusion that all observed $4\gamma$ events in the $\eta$ peak were likely
from the $\eta \to 3\pi^0$ decay they presented an
odd $\Gamma(\eta \to \pi^0\gamma\gamma)=(1.6\pm 4.7)\%\Gamma(
\eta\to\mathrm{neutrals})$ as their final result rather than just
an upper limit. 
We mention this curious
fact here only to show how great is sometimes the pressure of previously
published results --- a source of biases in high energy physics
experiments not always appreciated.

High $\gamma$ detection efficiency provided by heavy liquid bubble
chamber techniques was also used in $\eta \to \pi^0 \gamma \gamma$ 
decay studies. For example, for a xenon bubble chamber an average 
$\gamma$ detection efficiency was as high as 94\% \cite{12}. In this 
experiment,
the chamber was irradiated by a $2.34~\mathrm{GeV/c}\;\pi^+$ beam and
$\eta$ mesons were produced on quasi-free neutrons in the xenon nuclei
through the reaction $\pi^+ + n \to \eta +p$. Experiment \cite{13} also used
a xenon bubble chamber but $\pi^- + n \to \pi^- + n +\eta$ reaction to 
produce $\eta$-s by an incident $3.5~\mathrm{GeV/c} 
\;\pi^-$ beam. A variation of this technique, a hydrogen target inside a heavy
freon ($\mathrm{CF}_3\mathrm{Br}$) bubble chamber, was used in \cite{6,7}
with the $\eta$ meson production reaction $\pi^+ + p \to \pi^+ + 
p +\eta$. It was again emphasized in \cite{7} that the experimental 
difficulty in the $\eta \to \pi^0 \gamma \gamma$ decay studies comes
essentially from the $\eta \to 3\pi^0$ background.

Let us mention also two other bubble chamber experiments, for completeness.
In \cite{8}, $0.7\div 0.8~\mathrm{GeV/c} \; K^-$ beam at BNL 
Alternate Gradient Synchrotron (AGS) 
facility was used to produce $\eta$ mesons in a hydrogen bubble chamber 
via reaction $K^-+p\to\Lambda+\eta$. In \cite{4} the $\pi^++d\to p+p+\eta$
reaction in a bubble chamber filled with liquid deuterium was used,
$0.82~\mathrm{GeV/c} \; \pi^+$ beam provided also by AGS.

These bubble chamber experiments also produced controversial results. A real
breakthrough in the field occurred in 1981 with GAMS experiment at Serpukhov
\cite{14}.
The $\eta$ mesons were produced again via the charge exchange reaction
(\ref{eq:cher}). But now the incident $\pi^-$ beam momentum was 
$38~\mathrm{GeV/c}$. This incident energy increase dramatically improved
the $\eta\to 3\pi^0$ background suppression. In previous experiments
low energy ($\sim 1~\mathrm{GeV}$) $\eta$-s were used, which means relatively
high thresholds for $\gamma$-ray detection. So two softest $\gamma$-rays
could easily escape unnoticed, and this is one possible mechanism for
$\eta\to 3\pi^0$ decay to mimic the desired $\eta \to \pi^0 \gamma \gamma$
decay. With increasing $\eta$ energies, relative threshold for $\gamma$-ray 
detection gets lower,
and so does the $\eta\to 3\pi^0$ background. There is also
another benefit of higher energies: higher invariant mass resolution of the
detector for higher photon energies significantly improves suppression 
of the background.
An order of magnitude lower upper limit $\mathrm{Br}(\eta\to\pi^0\gamma
\gamma)<0.3\%$ (at 90\% confidence level) was reported by this experiment.

In the next run of the same GAMS experiment a lower $\pi^-$ momentum
$30~\mathrm{GeV/c}$ was used. But because of improvements in the 
detector and accelerator performance significantly higher statistics was
collected: about $6\times 10^5 \; \eta$ mesons produced via the charge 
exchange 
reaction (\ref{eq:cher}). At last the $\eta \to \pi^0 \gamma \gamma$ decay
was discovered with certainty \cite{15}. In the mass spectrum of the $\pi^0
\gamma\gamma$ system a narrow $\eta$ peak was observed over a smooth
background caused by the $\eta\to 3\pi^0$ decay. The statistical significance
of the effect was greater than seven standard deviations. The measured 
branching ratio was
$$\mathrm{Br}(\eta\to \pi^0 \gamma \gamma)=(9.5\pm 2.3)\times 10^{-4}.$$
\noindent After about two years the same experimental statistics was 
reanalyzed with improved reconstruction program for $\gamma$-rays from the
showers in the GAMS spectrometer \cite{16}. The improvement provided better
treatment  of overlapping showers and allowed to reduce the background
further. The new result was \cite{16,17}
\begin{equation}
\mathrm{Br}(\eta\to \pi^0 \gamma \gamma)=(7.1\pm 1.4)\times 10^{-4}.
\label{eq:br} \end{equation}

Now some words on history of theoretical studies of the 
$\eta\to \pi^0 \gamma \gamma$ decay.
To our knowledge, the first estimate of this decay rate
was given by Okubo and Sakita  in \cite{18}. They assumed that the 
following effective coupling
\begin{equation}
{\cal{L}}_{eff}=\xi F_{\mu \nu}F^{\mu \nu}\eta\pi^0 
\label{eq:leff} \end{equation}
\noindent was responsible for the decay. To estimate the value of the 
coupling constant $\xi$, they further assumed that the interaction
(\ref{eq:leff}) was also the source of the $\eta$-$\pi^0$ transition mass 
through the tadpole diagram which one obtains by contracting photon lines
in the $\eta\pi^0\gamma\gamma$ interaction vertex. The $\eta-\pi^0$ 
transition mass itself was estimated by using the unitary symmetry inspired
mass formula. This approach was hampered by the fact that 
the tadpole diagram is badly divergent and one needs some cut-off to 
calculate its strongly cut-off dependent value. By taking the cut-off at the
nucleon mass, they obtained $\Gamma(\eta\to \pi^0 \gamma \gamma)\sim
8~\mathrm{eV}$. But the authors admitted that this estimate may easily be 
wrong by an order of magnitude. For comparison, the experimental result   
(\ref{eq:br}) implies $\Gamma(\eta\to \pi^0 \gamma \gamma)=
0.84\pm 0.18~\mathrm{eV}$.

After some years, it was shown in \cite{19} that vector meson dominance
model (VDM) with trilinear meson interactions of the VVP and VPP type,
previously successfully applied to the $\eta\to\pi^+\pi^-\gamma$ and
$\eta\to\gamma\gamma$ decays \cite{20}, predicts too small ($\sim 10^{-3}$)
ratio of the $\eta\to \pi^0 \gamma \gamma$ and $\eta\to\gamma\gamma$ decay
widths, indicating $\Gamma(\eta\to \pi^0 \gamma \gamma)\sim 0.5~\mathrm{eV}$.
So it became clear that a large rate for the $\eta\to \pi^0 \gamma \gamma$
decay, which was reported by several experiments at that time, caused some
theoretical embarrassment. As a possible way out Singer
suggested in \cite{21} a rather strong quadrilinear VVPP meson interaction
which can be dominating in the $\eta\to \pi^0 \gamma \gamma$ decay.
The VDM model was further
examined by Oppo and Oneda in \cite{22}. It was confirmed that this model
predicts a small $\eta\to \pi^0 \gamma \gamma$ decay width in the 
$0.3$--$0.6~\mathrm{eV}$ range.
But the authors also indicated a possible additional
contribution coming from a scalar meson intermediate state. There was very
little experimental knowledge  about scalar mesons at that time (even nowadays
scalar mesons still remain the most experimentally unsettled question in the
low energy phenomenology). Using chiral symmetry arguments Gounaris \cite{23}
estimated scalar meson contribution and obtained
$\Gamma(\eta\to \pi^0 \gamma \gamma)=1.0\pm 0.2~\mathrm{eV}$ in excellent
agreement with the present experimental result (\ref{eq:br}). But this correct
experimental result then had years ahead to come. So the Gounaris' result
was considered as a failure of the theory rather than success. Some 
theoretical work followed \cite{24,25,26} which seemed successful in producing
large enough decay width by using essentially scalar meson contribution.
The new experimental situation \cite{14,15}, however, promptly depreciated all
such attempts and stopped them forever. 

Meanwhile it became increasingly evident that the chiral symmetry and its
spontaneous breaking play crucial role in the low energy hadron phenomenology.
This was culminated by creating a theoretical framework called chiral
perturbation theory \cite{27} where the degrees of freedom used are the
low-lying hadron states instead of useless in the low energy world (current) 
quarks and gluons. The situation here resembles \cite{28} theory of
superconductivity:
nobody questions the QED as underlying theory of all
electromagnetic phenomena, but it does not help much while dealing 
with superconductivity. The reason is that electrons and photons are not 
adequate degrees of freedom for the superconductive phase. The translational 
symmetry is spontaneously broken due to presence of atomic lattice.
The resulting Goldstone 
bosons (phonons) mediate the most important interactions. Electron gets 
a dynamical mass and becomes quite different from the electron of 
the bare QED Lagrangian. In the case of chiral perturbation theory (ChPT), 
the Goldstone bosons, associated with the spontaneous breaking of the chiral 
symmetry, are pseudoscalar mesons. So their interactions can be predicted  
to a great deal in the low energy limit. Therefore it is not surprising that
the last chapter of our theoretical history  is related to ChPT.

But at first let us mention some before-ChPT-age papers, for completeness.
It was shown in \cite{29,30} that, unlikely to $\eta\to\gamma\gamma$ decay,
baryon
loop contribution vanishes for the $\eta\to \pi^0 \gamma \gamma$ decay in
the chiral symmetry limit. Pion loop contribution turned out to be very
small \cite{30}. Failing to find any large contribution in the  
$\eta\to \pi^0 \gamma \gamma$ decay rate in otherwise successful chiral
model, the authors of \cite{30} suggested that something may be wrong with
the experiment itself and soon the GAMS experiment showed that they were 
correct. Right after this crucial experiment two papers appeared 
\cite{31,32} which calculated
the $\eta\to \pi^0 \gamma \gamma$ decay width by using (constituent) quark
loops to estimate various meson amplitudes. They obtained the correct 
magnitude of the decay width $\sim 1~\mathrm{eV}$.

In ChPT the $\eta\to \pi^0 \gamma \gamma$ decay was studied in \cite{33}.
This decay is rather peculiar from the point of view of the chiral
perturbation theory. In the momentum expansion, the leading $O(p^2)$ term is
absent because there is no direct coupling of photons to $\pi^0$ and
$\eta$. Due to the same reason the $O(p^4)$ tree contribution is also absent.
So for the $\eta\to \pi^0 \gamma \gamma$ decay ChPT series starts with
$O(p^4)$ 
one-loop contributions. But these loop contributions are also very small.
Usually dominant pion loops are suppressed here because they contain G-parity
violating $\eta\pi^+\pi^-\pi^0$ vertex. Kaon loops, although not
violating $G$-parity, also give negligible contribution because of large kaon 
mass.
It turned out that the main contribution comes from $O(p^6)$ counterterms
which are needed in ChPT to cancel various divergences. The coefficients of
these counterterms are not determined by the theory itself and should be fixed
either from experimental information or by assuming that they are saturated
by meson resonances (vector meson exchange giving the dominant
contribution). In \cite{33} the latter approach was adopted which gave
$\Gamma(\eta\to \pi^0 \gamma \gamma)\approx 0.18~\mathrm{eV}$ --- too small
compared to the experimental value (\ref{eq:br}). 
But, as was argued in \cite{33}, 
keeping momentum dependence in the vector meson propagators gives
an ``all-order'' estimate (that is $O(p^6)$ and higher) of the 
corresponding contribution of about $0.31~\mathrm{eV}$, in agreement 
with the old VDM prediction \cite{19,22}. Note that although the kaon loop 
amplitude alone is totally insignificant, its interference with the 
``all-order'' VDM amplitude gives still small but noticeable contribution 
\cite{34}. There are also scalar and tensor meson 
contributions (which signs cannot be unambiguously determined) and one-loop
contribution at $O(p^8)$ which is of the same 
order of magnitude
as the $O(p^4)$ contribution, because it does not violate G-parity.
Taking into account all these mechanisms, the final estimate of \cite{33} was
\begin{equation}
\Gamma(\eta\to \pi^0 \gamma \gamma)=0.42\pm 0.20~\mathrm{eV}.
\label{eq:brChPT} \end{equation} 
Some more details were given in \cite{35} but the result essentially remained
unchanged. Contributions of the C-odd axial vector resonances were considered
in \cite{36}. These contributions increase (\ref{eq:brChPT}) by about 10\%.
But this is still not enough to reconcile the ChPT result (\ref{eq:brChPT})
with the experiment. Therefore it was suggested in \cite{37} that the resonance
saturation assumption may not be valid for  $O(p^6)$ chiral Lagrangian 
(although it works quite well at $O(p^4)$). Some indications about validity
of this suggestion is provided by Bellucci and Bruno in \cite{38}. They
calculated the coefficients of the $O(p^6)$ counterterms in the framework
of the Extended Nambu Jona-Lasinio model (ENJL)  and obtained 
$$\Gamma(\eta\to \pi^0 \gamma \gamma)=0.58\pm 0.3~\mathrm{eV}.$$
\noindent It is 
higher than (\ref{eq:brChPT}) and is in agreement with experiment within one
standard deviation. However, the same coefficients when calculated by using
another chiral effective Lagrangian, obtained by a bosonization of the 
NJL model, led \cite{39} to small decay width of the order of 
$0.1~\mathrm{eV}$, while the $\gamma \gamma \to \pi^0 \pi^0$ experimental 
data were reproduced well. Another analysis \cite{40} within the ENJL model
framework also produced rather small $O(p^6)$ rate in the leading $1/N_c$
approximation
$$\Gamma(\eta\to\pi^0\gamma\gamma)=(0.27^{+0.18}_{-0.07})~\mathrm{eV}.$$
On the contrary, Nemoto et al. argued \cite{41} that the
ENJL model might be quite successful in explaining experimental 
$\eta\to \pi^0 \gamma \gamma$ width if one includes instanton induced
6-quark interactions, which lead to a strong $U_A(1)$ breaking and 
nonstandard $\eta$-$\eta^\prime$ mixing angle close to zero.

As we have seen, $\eta\to \pi^0 \gamma \gamma$ decay has dramatic, both 
experimental and theoretical, history. And the story is by no means
over yet. From the theoretical point of view, this decay provides
unique possibility to probe higher order effects in ChPT. But at present
these effects can not be calculated without model ambiguities. And the
situation will hardly change until more precise and detailed experimental
data appear. Such data about $\eta\to \pi^0 \gamma \gamma$ decay are also
highly desirable for understanding other interesting rare decays, such as
$K_L\to\pi^0 e^+ e^-$ \cite{42} (in which, it is believed, a signal of the
direct CP-violation can be observed) and $\eta\to\pi^0 l^+ l^-$ 
\cite{43,44,45} (which is a sensitive probe of C-conservation in the 
electromagnetic interaction of hadrons). Despite considerable experimental
effort, only one reliable measurement of the $\eta\to \pi^0 
\gamma \gamma$ decay rate exists at present. Therefore \cite{46},
``a re-measurement of the decay rate and a measurement of the decay 
distributions is certainly desirable''.

\section{Detector and experiment}
\begin{figure}[htb]
\begin{center}
\resizebox{0.65\textwidth}{!}{
  \includegraphics{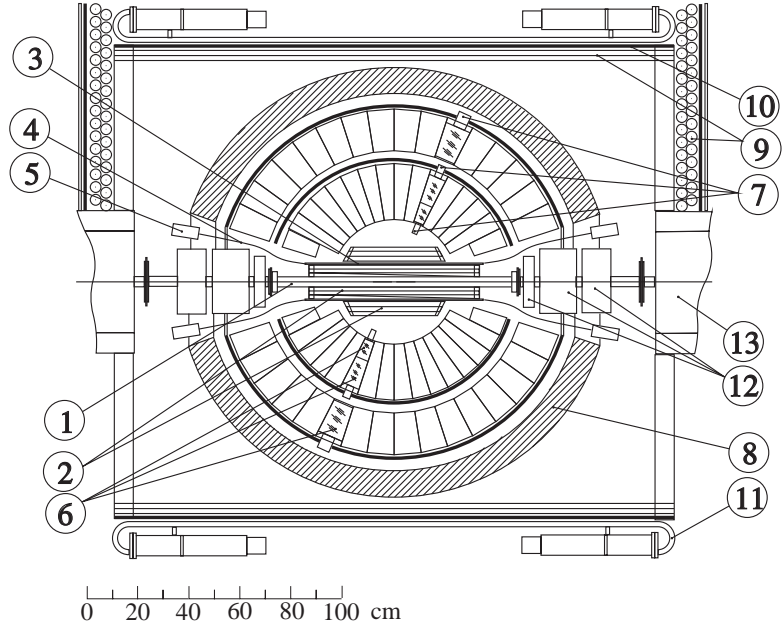}
}
\caption{ SND detector, section along the beams: (1) beam pipe,
         (2) drift chambers, (3) scintillation counter, (4) light guides,
         (5) PMTs, (6) NaI(Tl) crystals, (7) vacuum phototriodes,
         (8) iron absorber, (9) streamer tubes, (10) 1 cm iron plates,
         (11) scintillation counters, (12) and (13) elements of collider
         magnetic system.
}
\label{fig:SND}       
\end{center}
\end{figure}

The SND detector (Fig.\ref{fig:SND}) is a general purpose nonmagnetic 
detector which successfully operates at the VEPP-2M collider \cite{47} since
1995. The detailed description of this detector and some of its subsystems 
can be found elsewhere \cite{48,49,50}, and will be not repeated here. But
maybe it is worthwhile to mention some features of the SND calorimeter, 
because this subsystem is crucial in the analysis like one presented in 
this article. The SND ca\-lo\-ri\-me\-ter consists of 1630 NaI(Tl) crystals 
with a total weight of $3.6~{\rm tons}$. Crystals are assembled in 
three spherical shells with
solid angle coverage $\sim 90\%$ of $4\pi$. Each calorimeter layer 
includes 520--560 crystals of eight different shapes. The total thickness of
the calorimeter is $13.5 X_0$ (35~cm) of NaI(Tl). 

\begin{figure}[htb]
\resizebox{0.95\textwidth}{!}{
  \includegraphics{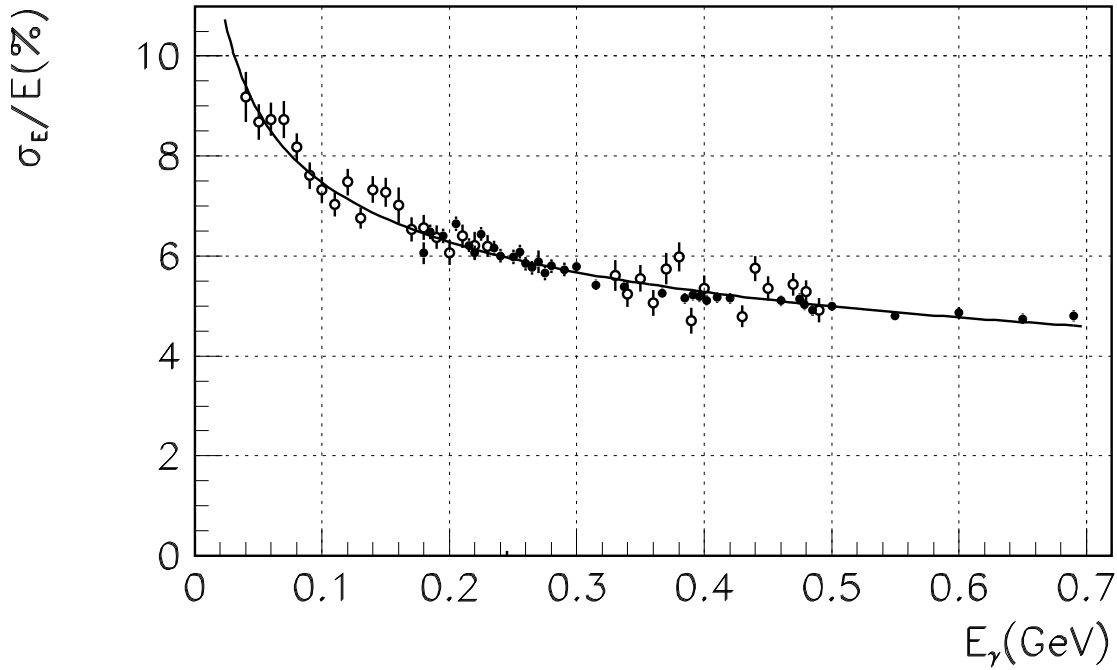}
}
\caption{Calorimeter energy resolution versus photon energy. 
         Energy resolution was measured by
         using $e^+e^- \rightarrow \gamma \gamma$ (dots) and
         $e^+e^- \rightarrow e^+e^- \gamma$ (circles) reactions.  
}
\label{fig:enr}       
\end{figure}

The calorimeter was calibrated by using cosmic muons \cite{51} and
$e^+e^-\to e^+e^-$ Bhabha events \cite{52}. The dependence of the calorimeter
energy resolution on photon energy is shown in Fig.\ref{fig:enr}.
This dependence can be fitted as:
$$\frac{\sigma_E}{E}(\%) = {4.2\% \over \sqrt[4]{E(\mathrm{GeV})}}.$$
Several effects influence the energy resolution, such as passive material
before and inside the calorimeter, leakage of shower energy through the 
calorimeter, electronics instability and calibration inaccuracy, light
collection nonuniformity over the crystal volume. Reasonably good agreement
between simulated and measured energy resolutions was achieved after all
these effects were duly taken into account in the simulation.
An average energy deposition for photons in the calorimeter is
about 93\% of their initial energy.

\begin{figure}[htb]
\resizebox{0.95\textwidth}{!}{
  \includegraphics{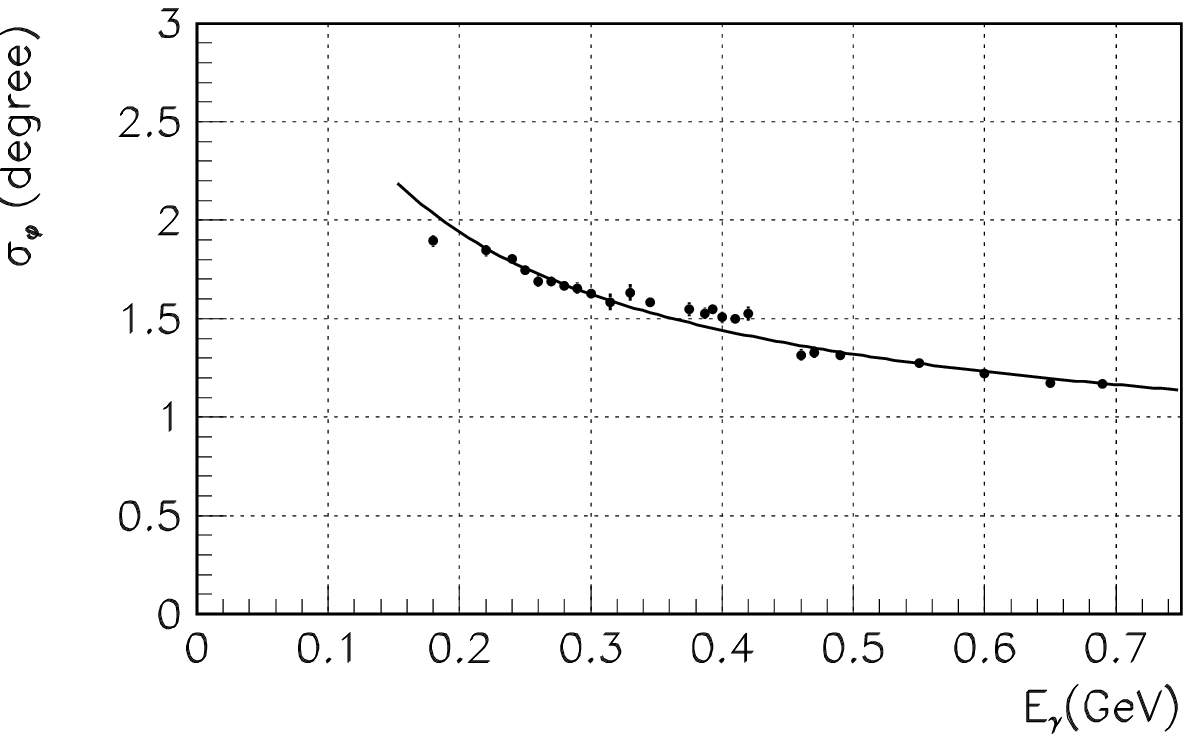}
}
\caption{Calorimeter angular resolution versus photon energy. 
}
\label{fig:anr}       
\end{figure}

High granularity of the calorimeter ensures rather good angular resolution 
for photons. A novel method for reconstruction of photon angles was 
suggested in \cite{53}. It is based on an empirical finding that the 
distribution function of energy deposition outside the cone with the angle
$\theta$ around the shower direction has the following form for 
electromagnetic showers in the SND calorimeter
(according to simulation) 
\begin{equation}
E(\theta) = \alpha \cdot exp(-~\sqrt[]{\theta/\beta}),
\label{eq:trprof} \end{equation}
\noindent where the $\alpha$ and $\beta$ parameters turned out 
to be practically
independent of the photon energy over the interval $50 \div 700$ MeV.
Fig.\ref{fig:anr} shows the calorimeter angular resolution as a function
of photon energy. This energy dependence can be approximated as
$$
\sigma_\varphi = {0.82^\circ \over \sqrt[]{E(\mathrm{GeV})}} \oplus 0.63^\circ.
$$

High granularity also enables to discriminate merged photons in the 
calorimeter by examining energy deposition profile of the corresponding 
electromagnetic shower in transverse direction. Merging of photons is
the main
mechanism for $\eta\to 3\pi^0$ decays to imitate the desired 
$\eta\to\pi^0\gamma\gamma$ process (another one is a loss of photons 
through openings in the calorimeter). So, for background
suppression, it is very important to recognize merged photons. In \cite{54}
a parameter was suggested to separate hadronic and electromagnetic showers
in the SND ca\-lo\-ri\-me\-ter by comparison of actual
 transverse energy profile of
the shower with the expected one
 (\ref{eq:trprof}) for a single photon. The 
same parameter also helps to discriminate merged photons.

The results presented in this article are based on the statistics collected
by SND in two experiments. The first one was carried out in the time period 
from February 1996 until January 1997. Seven successive scans were performed
in the center of mass energy range from 980 to 1044 MeV. Data were recorded 
at 14 different beam energy points. Collected statistics was 
almost tripled during  the next experiment in 1997--1998. It consisted of 
three scans at 16 energy points in the interval from 984 to
1060 MeV. The total integrated luminosity used in this work is equal to
$12.2~\mathrm{pb}^{-1}$ and corresponds to about $2\times 10^7\;\; \phi$ mesons
produced. The corresponding number of $\eta$ mesons from
$\phi\to\eta\gamma$ decay is $2.58\times 10^5$. The collider 
luminosity was measured using the process $e^+e^-\to 2\gamma$.  
Systematic error of the luminosity measurement 
estimated using $e^+e^-\to e^+e^-$ reaction is close to 3\%.

\section{Theoretical models for Monte Carlo simulation}
The general form of the $\eta(p)\to\pi^0(q)\gamma(k_1,\epsilon_1)
\gamma(k_2,\epsilon_2)$ transition amplitude 
as a function of particle momenta
and photon polarization vectors is dictated by 
gauge invariance, Bose symmetry, and CP-conservation \cite{45}:
$${\cal{M}}=\epsilon_{1\mu}\epsilon_{2\nu}\left [ A(x_1,x_2)T_{(1)}^{\mu\nu}
+\frac{B(x_1,x_2)}{m_\eta^2}T_{(2)}^{\mu\nu}\right ],$$
\noindent where $A(x_1,x_2)$ and $B(x_1,x_2)$ are 
Lorentz-invariant form factors, which are symmetric functions of dimensionless
variables 
$$x_1=\frac{p\cdot k_1}{m_\eta^2},\;\;\; x_2=\frac{p\cdot k_2}{m_\eta^2}.$$
\noindent Two independent gauge-invariant tensors are defined as
$$T_{(1)}^{\mu\nu}=k_2^\mu k_1^\nu-k_1\cdot k_2 g^{\mu \nu},$$
$$T_{(2)}^{\mu\nu}=k_1\cdot p ~k_2^\mu p^\nu+k_2\cdot p ~p^\mu k_1^\nu-
k_1\cdot k_2 ~p^\mu p^\nu -k_1\cdot p ~k_2\cdot p ~g^{\mu \nu}.$$

If CP is not conserved two more gauge-invariant structures
containing $\epsilon_{\mu\nu\lambda\sigma}$ (pseudo)tensor appear \cite{55}.
However we can safely neglect tiny effects associated with additional
form factors.

The decay width is given by the standard expression
\begin{equation}
d\Gamma=\frac{\delta(p-q-k_1-k_2)}{16~(2\pi)^5~m_\eta}
\sum\limits_{\epsilon_1 \epsilon_2}
|{\cal{M}}|^2~\frac{d\vec{q}}{E_\pi}
\frac{d\vec{k}_1}{E_1}\frac{d\vec{k}_2}{E_2}.
\label{eq:dGamma}\end{equation}
It is straightforward to perform summation over photon polarizations and obtain
$$\sum\limits_{\epsilon_1 \epsilon_2}
|{\cal{M}}|^2=\frac{1}{2}m_\eta^4\left\{ \left |A(x_1,x_2)+
\frac{1}{2}B(x_1,x_2)\right |^2
\left [2(x_1+x_2)+\frac{m_\pi^2}{m_\eta^2}-1\right]^2+ \right .$$
$$\left . \frac{1}{4}|B(x_1,x_2)|^2\left [ 4x_1x_2-2(x_1+x_2)-
\frac{m_\pi^2}{m_\eta^2}+1\right]^2 \right \}. $$
Then the decay width is given by the integral
$$\Gamma(\eta\to\pi^0\gamma\gamma)=\frac{1}{2!}~\frac{m_\eta}{64\pi^3}
\int\limits_{x_1^-}^{x_1^+}dx_1 \int\limits_{x_2^-}^{x_2^+}dx_2
~\sum\limits_{\epsilon_1 \epsilon_2}|{\cal{M}}|^2,$$
where the integration limits are 
$$x_1^-=0,\; x_1^+=\frac{1}{2}\left( 1-\frac{m_\pi^2}{m_\eta^2}
\right),\;
x_2^-=\frac{1}{2}\left (1-2x_1-\frac{m_\pi^2}{m_\eta^2}\right )$$
and
$$x_2^+=\frac{1}{2(1-2x_1)}\left (1-2x_1-\frac{m_\pi^2}{m_\eta^2}\right ).$$ 

We need some model to calculate $A(x_1,x_2)$ and $B(x_1,x_2)$ form factors
and proceed further. As was already mentioned 
in the introduction,
vector meson exchange $\eta\to V^*\gamma\to\pi^0\gamma\gamma$ gives the most
significant contribution. The corresponding form factors have the following
form \cite{44}
$$A_{VDM}(x_1,x_2)=$$
$$-\sum\limits_{V=\rho,\omega,\phi}g_{V\eta\gamma}g^*_{V\pi\gamma}
\left\{\frac{1-x_1}{1-2x_1-\frac{M_V^2}{m_\eta^2}}+
\frac{1-x_2}{1-2x_2-\frac{M_V^2}{m_\eta^2}}\right\},$$
$$B_{VDM}(x_1,x_2)=$$
$$\sum\limits_{V=\rho,\omega,\phi}g_{V\eta\gamma}g^*_{V\pi\gamma}
\left\{\frac{1}{1-2x_1-\frac{M_V^2}{m_\eta^2}}+
\frac{1}{1-2x_2-\frac{M_V^2}{m_\eta^2}}\right\}.$$
In the model considered in \cite{44}, which we choose as the base
for our Monte Carlo simulation, scalar meson exchange $\eta\to\pi^0 a^*_0\to
\pi^0\gamma\gamma$ contribution was also taken into account. 
If $a_0$ is considered as a point-like source of photon emission then
only $A(x_1,x_2)$ form factor acquires additional contribution from this 
mechanism
$$A_{a_0}(x_1,x_2)=\frac{g_{a\eta\pi}g_{a\gamma\gamma}}{m_a^2-m_\eta^2
\left [ 2(x_1+x_2)+\frac{m_\pi^2}{m_\eta^2}-1\right ]}.$$
For the coupling constants we take their central values from Ref.\cite{44}, 
where they were extracted from the experimental information on various
decay widths. The relative sign of the VDM and $a_0$ exchange amplitudes 
cannot be determined this way. 
We have chosen constructive
interference between them because the experimental width is larger than 
VDM prediction.

The following resulting values were used in the simulation:
$$m_\eta^2g_{\rho\eta\gamma}g^*_{\rho\pi\gamma}=0.0495,\;\;\;
m_\eta^2g_{\omega\eta\gamma}g^*_{\omega\pi\gamma}=0.0294, $$
$$m_\eta^2g_{\phi\eta\gamma}g^*_{\phi\pi\gamma}=0.00267,\;\;\;
g_{a\eta\pi}g_{a\gamma\gamma}=0.0132.$$
As a check of the simulation program,
we reproduced the result of Ref.\cite{44} for the decay
width $\Gamma(\eta\to\pi^0\gamma\gamma)_{VDM+a_0}=0.36~\mathrm{eV}$.

Equation (\ref{eq:dGamma}) can be rewritten in a form more convenient for 
simulation. The $\delta$-function (or energy-momentum conservation) enables us
to perform a trivial
integration over $d\vec{q}$ and $d\cos{\theta_2}$ (where $\theta_2$
is the angle between $\vec{k}_1$ and $\vec{k}_2$ 3-vectors) and
get in the $\eta$-meson rest frame:
\begin{equation}
d\Gamma=\frac{m_\eta}{64\pi^3}
\sum\limits_{\epsilon_1 \epsilon_2}
|{\cal{M}}|^2~dx_1dx_2
\frac{d\cos{\theta_1}}{2}\frac{d\varphi_1}{2\pi}\frac{d\varphi_2}{2\pi}\; .
\label{eq:dGamma1}\end{equation}
$x_1,\;x_2$ which were defined earlier, in the $\eta$-meson rest frame
coincide with photon energies divided by $\eta$-meson mass. $\theta_1$ and
$\varphi_1$ are the first photon polar and azimuthal angles in the
$\eta$ meson rest frame and $\varphi_2$ is the second photon azimuthal angle
in the plane perpendicular to the first photon momentum $\vec{k}_1$. In fact
the integrand in (\ref{eq:dGamma1}) after taking into account the
energy-momentum conservation depends only on $x_1$ and $x_2$.
It is convenient, however,
to introduce another angular variable
instead of $x_2$ in the following way. Let $S^*$ be a rest frame of
the system consisting of $\pi^0$ and the second photon.
Of course this system moves in the direction opposite to the $\vec{k}_1$
if initial $\eta$ decays at rest.
For the energy $E_*$ and invariant mass $m_*$ of the $\pi^0\gamma$ system
we get easily
\begin{equation}
E_*=m_\eta(1-x_1), \;\;\; m_*=m_\eta\sqrt{1-2x_1}.
\label{eq:e*m*}\end{equation}
Therefore the Lorentz factor of this system is
\begin{equation}
\gamma=\frac{E_*}{m_*}=\frac{1-x_1}{\sqrt{1-2x_1}}.
\label{eq:gamma}\end{equation}
The energy of the second photon $E^*_2$ in the $S^*$ reference frame 
is uniquely determined by $m_*$ and hence by $x_1$:
\begin{equation}
x_2^*=\frac{E_2^*}{m_\eta}=\frac{m^{*2}-m_\pi^2}{2m_\eta m_*}=
\frac{1-2x_1-\frac{m_\pi^2}{m_\eta^2}}{2\sqrt{1-2x_1}},
\label{eq:x2*}\end{equation}
but its direction is arbitrary.
Let us denote the polar and azimuthal angles
($\theta^*_2$ and $\varphi^*_2$) of 
the second photon in $S^*$ as $\varphi^*_2=
\varphi_2$ and $\theta^*_2$ --- as an angle between the velocity 3-vector
of the reference frame $S^*$
and the direction of the second photon. The Lorentz transformation
relates $x_2$ and $\cos{\theta^*_2}$ in the following way
$$x_2=\gamma x_2^*(1+\beta \cos{\theta^*_2}),\;\;\; 
\beta=\frac{\sqrt{\gamma^2-1}}{\gamma}.$$ 
Therefore we can rewrite (\ref{eq:dGamma1}) as 
\begin{equation}
d\Gamma=\frac{m_\eta}{64\pi^3}f(x_1)
\sum\limits_{\epsilon_1 \epsilon_2}
|{\cal{M}}|^2~dx_1
\frac{d\cos{\theta_1}}{2}\frac{d\varphi_1}{2\pi}\frac{d\cos{\theta^*_2}}{2}
\frac{d\varphi^*_2}{2\pi}\; ,
\label{eq:dGamma*}\end{equation}
where
$$f(x_1)=\frac{dx_2}{d\cos{\theta^*_2}}=\beta\gamma x^*_2=\sqrt{\gamma^2-1}
~x^*_2.$$

Having equation (\ref{eq:dGamma*}) at hand, we can generate $\eta\to\pi^0
\gamma\gamma$ decay in the $\eta$ meson rest frame by the following algorithm:
\begin{itemize}
\item[$\bullet$] generate first photon normalized energy $x_1$ as a random 
number uniformly distributed from $0$ to $x_1^+$. Calculate 
$(\pi^0,\gamma)$-system energy, mass and the Lorentz factor according to 
equations (\ref{eq:e*m*}) and  (\ref{eq:gamma}).
\item[$\bullet$] generate a random number $\varphi_*$ uniformly distributed 
in the interval $[0,2\pi]$ and take it as the azimuthal angle of the $S^*$ 
velocity vector in the $\eta$ meson rest frame, which will be called 
laboratory frame below. Generate another uniform random number in the 
interval $[-1,1]$ and take it as a $\cos{\theta_*},\;\theta_*$ being the 
polar angle of the $S^*$ velocity vector in the laboratory frame. This 
defines the unit vector
$\vec{n}=(\sin{\theta_*}\cos{\varphi_*}, \sin{\theta_*}\sin{\varphi_*},
\cos{\theta_*})$ along $S^*$ velocity.
\item[$\bullet$] now the first photon 4-momentum in the laboratory frame can 
be constructed: $E_1=m_\eta x_1,\; \vec{k}_1=-E_1\vec{n}$. 
\item[$\bullet$] generate in the analogous manner $\varphi^*_2$ and 
$\cos{\theta^*_2}$,
construct the unit vector $\vec{n}^*$ along the second
photon velocity in the $S^*$ frame: $$\vec{n}^*=(\sin{\theta^*_2}
\cos{\varphi^*_2},\sin{\theta^*_2}\sin{\varphi^*_2}, \cos{\theta^*_2}).$$
\item[$\bullet$] calculate $x^*_2$ according to (\ref{eq:x2*}) and construct
the second photon 4-momentum in the $S^*$ frame: $E^*_2=x^*_2m_\eta,\;$
$\vec{k}^*_2=E^*_2\vec{n}^*.$
\item[$\bullet$] construct the $\pi^0$ meson 4-momentum in the $S^*$ frame:
$E^*_\pi=m_*-E^*_2,\; \vec{q}_\pi^{\, *}=-\vec{k}^*_2.$
\item[$\bullet$] construct the $\pi^0$ meson and second photon 4-momenta in 
the laboratory frame from their 4-momenta in the frame $S^*$ by using 
appropriate Lorentz transformations.
\item[$\bullet$] for generated $x_1$ and $x_2$, calculate
$$z=f(x_1)\sum\limits_{\epsilon_1 \epsilon_2}|{\cal{M}}|^2.$$
\item[$\bullet$] generate a random number $z_R$ uniformly distributed in the
interval from 0 to $z_{max}$, where $z_{max}$ is some number majoring $z$. 
That is for all $\eta\to\pi^0\gamma\gamma$ decay configurations (for all 
acceptable values of $x_1$ and $x_2$) we should have $z<z_{max}$. In our
program we had used $z_{max}=1.8\times 10^{-5}.$
\item[$\bullet$] if $z\ge z_R$, accept the event, that is the generated 
4-momenta of the $\pi^0$ meson and both photons. Otherwise repeat the whole
procedure.   
\end{itemize} 
Note that equation (\ref{eq:dGamma*}) indicates that the decay width can be 
calculated by using the above described Monte Carlo algorithm according to
$$\Gamma(\eta\to\pi^0\gamma\gamma)=\frac{1}{2!}~\frac{m_\eta}{64\pi^3}
\left <f(x_1)\sum\limits_{\epsilon_1 \epsilon_2}|{\cal{M}}|^2\right >
(x_1^+-x_1^-),$$
where $<\ldots>$ stands for the mean value. As another check of the 
program, we had reproduced the result of \cite{44} for the width by this
method also.

To understand whether model dependence of the $\eta\to\pi^0\gamma\gamma$
decay matrix element can bring a dangerous systematics in our 
experimental study of this decay, we performed another Monte Carlo simulation
of this process using significantly different form factors of the quark-box 
model from \cite{45}. These form factors are reproduced below (in the units 
of $\mathrm{GeV}^{-2}$): 
$$\left [2(x_1+x_2)+\frac{m_\pi^2}{m_\eta^2}-1\right]A_{QB}(x_1,x_2)\approx
$$ $$-0.616+2.14(x_1+x_2)-2.509(x_1^2+x_2^2)-4.184x_1x_2+ $$ $$
1.5896(x_1^3+x_2^3)+2.936x_1x_2(x_1+x_2),$$
$$B_{QB}(x_1,x_2)\approx -0.866+1.674(x_1+x_2)-3.26(x_1^2+x_2^2)-$$ 
$$1.781x_1x_2+2.37(x_1^3+x_2^3)+1.089x_1x_2(x_1+x_2).$$
For the quark-box model, we had used $z_{max}=4.37\times 10^{-5}$ as majoring
value in our Monte Carlo program. The prediction of this model 
for the decay width is about two
times higher than the VDM prediction and is closer to the present 
experimental result. But this model is also more ambiguous. Its predictions
strongly depend on the assumed values for the $u$ and $d$ constituent quark 
masses (the above given form factors correspond to $m_u=m_d=300~\mathrm{MeV}$).
It was proven that in the chiral symmetry limit baryon loop contribution
strictly vanishes \cite{29,30}. In order to check whether quark-box
diagaram contribution is indeed dominant we must consider other quark-loop
diagrams also in the framework of some self-consistent model (for example 
ENJL model) with realistically broken chiral symmetry. Therefore we took
VDM as a basis for our Monte Carlo simulation and used the quark-box model 
only for systematic error estimation.

\section{Data analysis and results}
A primary selection of the candidate events for the process
\begin{equation}
e^+e^-\to\phi\to\eta\gamma;\;\eta\to\pi^0\gamma\gamma
\label{eq:etpgg}\end{equation} 
was done according to following criteria:
\begin{itemize}
\item[$\bullet$] an event must have no charged particles and
contain exactly 5 photons in the calorimeter; 
\item[$\bullet$] the azimuthal angle of any final photon lies within the 
interval $27^\circ<\theta<153^\circ$;
\item[$\bullet$] the energy of the next to the most energetic photon is less
than $0.8E$, $E$ being a beam energy; 
\item[$\bullet$] the total
energy deposition of final photons is in the range $0.8 \le E_{tot}/2E 
\le 1.1$;
\item[$\bullet$] the normalized full momentum of the event 
($P_{tot}/2E$) is less than 0.1;
\end{itemize}
The third condition is against QED background $e^+e^-\to 2\gamma$, with extra 
photons originating either from the beam background or splitting 
of electromagnetic showers in the calorimeter. The two latter conditions 
eliminate the main part of the  background from the $\phi \to 
K_LK_S$ decay.  

For the events which passed the primary selection criteria a kinematic fit
was performed assuming energy-momentum conservation and the presence of a
$\pi^0$ meson in an intermediate state. $\chi^2$ of this fit ($\chi_\pi$)
can be used to suppress the most dangerous background from
\begin{equation}
e^+e^-\to \eta\gamma;\;\;\; \eta\to 3\pi^0\;.
\label{eq:eta3pi} \end{equation}
In particular the $\chi_\pi<10$ condition being used in our event 
selection procedure reduces the background (\ref{eq:eta3pi})
by a factor of two.
Another $\sim20\%$ reduction of this background was achieved by demanding that
any photon pair not involving photons from the 
$\pi^0$ meson found by the kinematic fit has an invariant mass outside
$m_{\pi^0}\pm 30~\mathrm{MeV}$ interval.

Another significant background left after the selection criteria described 
above comes from the process
\begin{equation}
e^+e^-\to \pi^0\omega\to\pi^0\pi^0\gamma\;.
\label{eq:ompi} \end{equation}
This background is only slightly ($\sim 15\%$) reduced by the $\chi_\pi<10$
cut but is suppressed by an order of magnitude after the invariant
masses of photon pairs
are restricted as described above to exclude the presence of a second 
$\pi^0$ meson.

To further reduce this background, special kinematic fit was applied to the 
selected events assuming $\omega \pi^0$ intermediate state. Only the events
for which this kinematic fit fails were selected for further analysis. 
 
One more background source is the $e^+e^-\to 3\gamma$ reaction (both pure QED
and $\phi\to\eta\gamma, \eta\to 2\gamma$
decay) with extra background photons. These extra photons are usually soft.
To suppress this background, a $3\gamma$-kinematic 
fit was performed under assumption that two photons in the event were
spurious. 
All $3\gamma$-combinations were examined and one with the best $\chi^2$
(denoted as $\chi_{3\gamma}$) was selected. The $e^+e^-\to 3\gamma$
background is efficiently
suppressed by the condition  $\chi_{3\gamma}>20$, while the process 
(\ref{eq:etpgg}) itself is only marginally effected.
 
\begin{figure}[htb]
\begin{center}
\resizebox{0.65\textwidth}{!}{
  \includegraphics{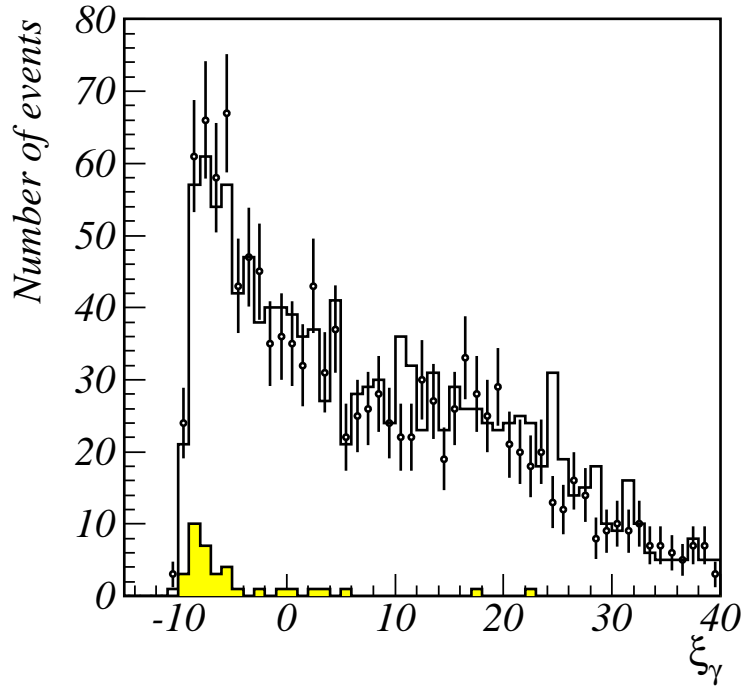}
}
\caption{$\xi_\gamma$ distributions. Histogram --- simulated
background (\ref{eq:eta3pi}) normalized to experimental statistics.
Points with error bars --- experiment. Shaded histogram --- expected
$\eta\to \pi^0\gamma\gamma$ signal according to simulation of the process 
(\ref{eq:etpgg}). }
\label{fig:xinm}
\end{center}       
\end{figure}

As was already mentioned, transverse energy profile of the shower enables us to
recognize merged photons and additionally suppress the  background 
(\ref{eq:eta3pi}) . The corresponding parameter $\xi_\gamma$ is described in 
\cite{54}. Fig.\ref{fig:xinm} shows distributions over the parameter 
$\xi_\gamma$ for simulated signal events (\ref{eq:etpgg}) and background 
(\ref{eq:eta3pi}). The efficiency of this parameter in the background 
suppression is clearly demonstrated by this figure. We require $\xi_\gamma<-5$ 
which reduces background (\ref{eq:eta3pi}) by about six 
times.

Fig.\ref{fig:xinm} also shows that almost all background remaining
after our cuts
comes from the process (\ref{eq:eta3pi}). In particular the background
from the reaction (\ref{eq:ompi}) seems to become insignificant after
application of described above selection criteria.
But unfortunately it is very 
difficult to suppress the background (\ref{eq:eta3pi}) any further without 
significant
improvement of the detector performance (angular and energy resolution for 
photons). So the only option left for us is to subtract the remaining
background (\ref{eq:eta3pi}) relying upon the Monte Carlo simulation.

\begin{figure}[htb]
\begin{center}
\resizebox{0.65\textwidth}{!}{
  \includegraphics{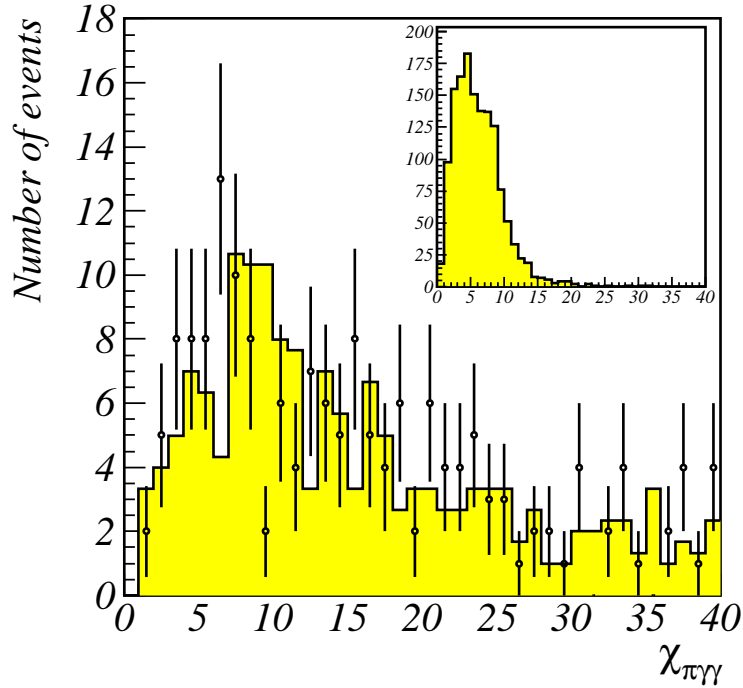}
}
\caption{$\chi_{\pi\gamma\gamma}$ distribution. Points with error bars --- 
experiment. Shaded histogram corresponds to the  
simulated background (\ref{eq:eta3pi}), normalized to experimental 
statistics. Insertion shows the distribution for simulated 
$\eta\to \pi^0\gamma\gamma$ events.}
\label{fig:zxi2}
\end{center}       
\end{figure}

For this purpose, special kinematic fit was performed for selected
events. It was assumed in the fit that four of five photons in each
event are from the $\eta\to\pi^0\gamma\gamma$ decay. Of course there are 
many possibilities how to divide five photons into the recoil photon from
the $\phi\to\eta\gamma$ decay, two photons from the subsequent $\eta\to\pi^0
\gamma\gamma$ decay, and two photons from the $\pi^0\to\gamma\gamma$ decay. The
best combination was selected for each event during the fit. $\chi^2$ of the 
fit ($\chi_{\pi\gamma\gamma}$) is small for desired events from 
(\ref{eq:etpgg}), while background (\ref{eq:eta3pi}) has wider distribution.
Fig.\ref{fig:zxi2} shows $\chi_{\pi\gamma\gamma}$ distribution for 170
experimental events left after the above described cuts with one more 
condition $\chi_{\pi\gamma\gamma}<40.$ The simulated background from the 
process (\ref{eq:eta3pi}) has similar distribution over this parameter.

Signal and background composition of our data sample was estimated by using 
binned maximum likelihood method incorporating the finiteness of the Monte
Carlo statistics \cite{56}. Using HMCMLL routine from the HBOOK package
\cite{57}, the experimental histogram was fitted by the sum
of two Monte Carlo histograms for the signal (\ref{eq:etpgg}) and background
(\ref{eq:eta3pi}). As a result of this fit, the signal events fraction in the
experimental data, depicted in Fig.\ref{fig:zxi2}, was found to be $(4.1\pm
7.6)\%$. This corresponds to the total number of signal events of
the process (\ref{eq:etpgg})  $N_S=7.0^{+12.9}_{-6.5}$, where the asymmetric 
errors correspond to the 68.27\% confidence interval 
determined according to Table 10 from \cite{58}. Number of background events
from this fit $N_B=163\pm 19$ is in a good agreement with the prediction of 
simulation $158.3\pm 11.5$. 

To transform $N_S$
into the branching ratio, we need a detection efficiency for signal 
events. This efficiency determined from simulation
of the effect with VDM matrix element is $\epsilon=
(14.1\pm 0.4)\%$. The quark-box matrix element gives almost the same value
$\epsilon=(14.8\pm 0.4)\%$. Therefore the systematic error coming from
model ambiguity is negligible in comparison with the inaccuracy in the
background subtraction. The Monte Carlo detection efficiency was corrected
for a loss of events due to additional beam background photons causing
signal events to be rejected by the selection cut requiring exactly 5 photons
in an event.
This effect not modelled
in the simulation was as large as 5\% in the first SND experiment and
8\% in the second one with a twice larger statistics. Thus as an
average corrected detection efficiency for signal events
we took $\epsilon=(13.1\pm 0.8)\%$ with the model  ambiguity also taken
into account.

The number of $\eta$ mesons in
our initial data sample is $2.58\times 10^5$ 
with a 3\% systematic error due to limited accuracy of the luminosity
determination. Given above values for $N_S$ and $\epsilon$ we get the
branching ratio 
\begin{equation}
Br(\eta \to \pi^0 \gamma \gamma)=(2.1^{+3.8}_{-1.9})\times 10^{-4}.
\label{eq:res1} \end{equation}
Since the error is very large only an upper limit of the branching ratio
can be determined from our data: 
\begin{equation}
Br(\eta \to \pi^0 \gamma \gamma)<8.4\times 10^{-4} \, \mbox(90\%\, C.L.),
\label{eq:res} \end{equation}
\noindent where we have again used Table 10 from \cite{58}, according to which
the measured mean of $0.54\sigma$ implies 90\% C.L. interval $\mu<2.18\sigma$ 
for the mean $\mu$ of a Gaussian constrained to be non-negative.

The systematic error caused mainly by inaccuracy of the simulated 
$\xi_\gamma$-distribution was estimated to be less than 20\%
which is negligible in comparison with the statistical error.

\iftranslate
To check consistency  of this particular background
subtraction procedure we
have tried another approach.
The $\chi_{\pi\gamma\gamma}<40$ region in the experimental histogram
of Fig.\ref{fig:zxi2} was divided into two bands: 1 --- with
$\chi_{\pi\gamma\gamma}\le\chi_R$ and 2 --- with $\chi_R<\chi_{\pi\gamma
\gamma}<40.$ Let $N_1$ and $N_2$ be the numbers of experimental events in
the corresponding bands. Then we can write
\begin{equation}
N_1=N_Sk_S+N_B(1-k_B),\;\;\; N_2=N_S(1-k_S)+N_Bk_B,
\label{eq:N12} \end{equation}
\noindent where $N_S,\;N_B$ are the numbers of signal and background events
in the 
whole region $\chi_{\pi\gamma\gamma}<40$, $k_s$ is the fraction of signal
events from this region belonging to the band 1, and $k_B$ is the 
fraction of the background events belonging to the band 2. The values of
$k_S$ and  $k_B$ were determined by simulation. After that  $N_S$ and $N_B$
can be easily calculated from
(\ref{eq:N12}). The results of this 
procedure for various choices of $\chi_R$ are listed in the table 
\ref{tab:1}.
\begin{table}[htb]
\begin{center}
\caption{Background subtraction results}
\label{tab:1}       
\begin{tabular}{ccccc}
\hline\noalign{\smallskip}
$\chi_R$ & 5 & 10 & 15 & 20 \\
\noalign{\smallskip}\hline\noalign{\smallskip}
$k_S$ & $0.44\pm 0.02$ & $0.88\pm 0.03$ & $0.98\pm 0.04$  & $0.99\pm 0.04$ \\
$k_B$ & $0.88\pm 0.06$ & $0.61\pm 0.05$ & $0.41\pm 0.03$ & $0.28\pm 0.03$ \\
$N_1$ & $23\pm 4.8$ & $64\pm 8$ & $92\pm 9.6$  & $117\pm 10.8$ \\
$N_2$ & $147\pm 12.1$ & $106\pm 10.3$ & $78\pm 8.8$  & $53\pm 7.3$ \\
$N_S$ & $8.1\pm 33.4$ & $-4.7\pm 22.0$ & $-21.3\pm 22.4$ & $-20.0\pm 31.0$  \\
$N_B$ & $161.9\pm 35.6$ & $174.7\pm 25.8$ & $191.3\pm 26.7$ 
& $190.0\pm 34.2$ \\
\noalign{\smallskip}\hline
\end{tabular}
\end{center}
\end{table}

Note that, according to statistically more powerfull previous method,
the number
of the $\eta\to\pi^0\gamma\gamma$ events is $7.0^{+12.2}_{-6.7}$ and 
simulation
predicts $158.3\pm 11.5$ background events. These numbers are consistent with 
$N_S$ and $N_B$ from the Table \ref{tab:1}.  

\begin{figure}[htb]
\begin{center}
\resizebox{0.65\textwidth}{!}{
  \includegraphics{egr.eps}
}
\caption{Recoil mass of the most energetic photon. Points with error bars  
correspond to experimental events. Shaded histogram --- estimated background 
from the process (\ref{eq:eta3pi}). }
\label{fig:egr}
\end{center}       
\end{figure}
 
Fig.\ref{fig:egr} shows recoil mass of the most energetic photon for 
experimental events with $\chi_{\pi\gamma\gamma}<10$.
$\eta$-meson signal is clearly seen, but this signal is still vastly populated
by the background (\ref{eq:eta3pi}). As our last resource to extract the
$\eta\to\pi^0\gamma\gamma$ decay signal from the background, let us try to use
peculiarities of the pion energy spectrum in this decay. In the $\eta$ meson
rest frame, this energy spectrum is moderately peaked at high energies. This
is true for both VDM and quark-box models, although in the latter case the 
effect is somewhat less prominent. Fig.\ref{fig:xf} shows distributions over 
the parameter $x=1-x_1-x_2$, where $x_1$ and $x_2$ are normalized over $m_\eta$
photon energies in the $\eta$ meson rest frame. So $x$ is in fact the 
normalized pion energy in this frame. The hole in the distribution near $x=
0.5$ is caused by our selection criteria, which are against the presence of
the second $\pi^0$ in an event. Indeed, the photon pair has invariant mass
$m^2_{\gamma\gamma}=m_\pi^2+m_\eta^2(1-2x)$. So, demanding single 
$\pi^0$ in an event we exclude some region near $x=0.5$.

\begin{figure}[htb]
\begin{center}
\resizebox{0.65\textwidth}{!}{
  \includegraphics{xf.eps}
}
\caption{Pion energy spectrum in the $\eta$ meson rest frame. Energy is 
in the units of the $\eta$-meson mass. Points with 
error bars --- experiment, shaded histogram --- estimated background 
from the process (\ref{eq:eta3pi}). Insertion shows corresponding
distribution for simulated  $\eta\to \pi^0\gamma\gamma$ events.}
\label{fig:xf}
\end{center}       
\end{figure}

Fig.\ref{fig:xf} shows the spectra without $\chi_{\pi\gamma
\gamma}$ cut, although all other selection criteria described above were 
applied. But higher $x$, 
which we are interested in now, means softer photons. So the probability of 
superimposed  beam background photons increases.
To exclude this background, we demand 
$\chi_{\pi\gamma\gamma}<10$. Then only 6 experimental events are left in the
$x>0.5$ region, while the simulation predicts $3.3\pm 1.0$ background events
and $2.8\pm 0.6$ signal events, for the branching ratio (\ref{eq:br}). 
So what is observed is consistent with what is expected, but again only an
upper limit can be established with certainty. 
The detection efficiency for signal events according to Monte Carlo
is (decreased by 7\% to correct beam background related systematics as 
explained earlier) 1.5\% for VDM
and 1.3\% for quark-box model. Then our 6 observed event with a mean background
of $3.3$ indicate the following 68.27\% C.L. interval, according to
Feldman and Cousins table 2 \cite{58} 
\begin{equation}
Br(\eta\to\pi^0\gamma\gamma)=(7.0^{+7.9}_{-5.4})\times 10^{-4}.
\label{eq:res2} \end{equation}
Systematic errors are potentially large. We do not 
scrutinize them because
the signal is not statistically significant anyway.  
\fi

\section{Final remarks}
Our investigation once more clearly demonstrates experimental difficulties
in $\eta\to\pi^0\gamma\gamma$ decay studies caused mainly by the background
(\ref{eq:eta3pi}).
Our analysis yields the value of the $\eta\to\pi^0\gamma\gamma$
branching ratio (with systematic error not grater than 20\%)
$Br(\eta\to\pi^0\gamma\gamma)=(2.1^{+3.8}_{-1.9})\times 10^{-4}$
which is consistent with
a more accurate result of GAMS \cite{16,17}:
$Br(\eta\to\pi^0\gamma\gamma)=(7.1\pm 1.4)\times 10^{-4}$,
but shows an urgent need of new high precision measurements.
As a final result we present an upper limit:
$$Br(\eta \to \pi^0 \gamma \gamma)<8.4\times 10^{-4}\, \mbox{at 90\% C.L.,}$$
leaving open the question whether ChPT has problems with this decay.
Nevertheless, we think that the investigation presented in this paper
demonstrates feasibility of such studies in future high statistics collider
experiments. Let us remind in this respect that
$\eta$ mesons will be copiously produced at $\phi$-factories,
for example,
about $10^8\;$ $\phi\to\eta\gamma$ decays per year are expected at
DA$\Phi$NE \cite{59}. We hope that this article will stimulate future 
investigations at $\phi$-factories and will be useful in such studies.

\section*{acknowledgement}
This work is supported in part by Russian Fund for Basic Researches, grant
No. 00-15-96802.

\end{document}